\documentclass[12pt]{article}
\usepackage{epsfig}
\usepackage{float}
\begin{document}
\newcommand{\beq}{\begin{equation}}
\newcommand{\bx}{\mathbf x}
\newcommand{\eeq}{\end{equation}}

\begin{center}
{\bf \Large A Modified Schmidl-Cox OFDM Timing Detector}\\

\vspace{0.1in}
{\large Stephen G. Wilson, Rui Shang\\
Department of Electrical and Computer Engineering\\
University of Virginia\\
Charlottesville, VA  22904}

\end{center}

\section{Abstract}

We describe a simple modification of the Schmidl-Cox detector for establishing timing in OFDM transmissions that stabilizes performance in transitions from no-signal
to signal, or vice-versa.  Moreover, the proposed modification scales the detector's
metric between 0 and 1 for all scenarios, simplifying threshold setting, and improves timing detector SNR.

\section{Introduction}
Schmidl and Cox \cite{Schmidl} proposed a simple two-symbol preamble for establishing frame timing and frequency offset estimation in OFDM.  The first symbol provides a time-domain sequence whose second-half samples are identical to those in the first half.  This is imposed in OFDM by zeroing alternate frequency-domain variables.\footnote{More broadly, this procedure could be applied to locate any non-OFDM pattern that has such repetition in time.}  Normally placed at the beginning of an OFDM burst, this symbol alllows detection of start-of-frame (SOF) as described below.  This same detection procedure also provides coarse estimation of the
beginning of OFDM symbols to follow.

Though not under consideration here,  Schmidl and Cox also showed how frequency estimation can
be performed from this first preamble symbol, to within an ambiguous interval corresponding
to the OFDM subcarrier spacing.  A second preamble symbol is used to resolve ambiguity of
frequency--see \cite{Schmidl} for details.

In related work, Minn et al \cite{Minn} published a modification that sharpens the SOF peak,  by modifying the S-C symbol
construction. Other preamble-based designs are reported in \cite{Park, Liu, Sreedhar}.  Some of this work has focused on maximum likelihood estimation of timing and/or frequency offset using the cyclic prefix property or the preamble structure itself, \cite{Sandell, Shi}.  These could possibly be used
to refine over time S-C estimates obtained from the two preamble symbols.

In the following, we denote the FFT size in an OFDM implementation by $N$, and  let $s_n$ denote the time-domain sequence at the output of the $N$-point IFFT in OFDM transmitter.  This sequence is acted upon by a  multipath channel , then additive white Gaussian noise to produce the complex baseband sequence at the receiver
\beq
r_n = \sum_{m=1}^D h_m s_{n-m} + n_n~,
\eeq
where $D$ is the anticipated multipath channel duration in samples.  Typically a cyclic-prefix is added to the IFFT output, having length $N_{CP} > D$, whose removal at the receiver
allows easy frequency-domain equalization of the channel.

Letting $L=N/2$ denote half the symbol duration (wthout CP), the S-C procedure \cite{Schmidl}
first defines
\beq
P(n) = \sum_{m=0}^{L-1} r_{n+m}^* r_{n+m+L}~, 
\eeq
a sliding lag-$L$ correlation of the received sequence computed over $L$ samples,
and 
\beq
R(n)=\sum_{m=0}^{L-1} |r_{n+m+L}|^2
\eeq
which is energy measured over $L$ contiguous samples.  (Both are non-causal as defined.)  Then the ratio
\beq
M(n) = \frac{|P(n)|^2}{R^2 (n)}
\eeq
is formed whose peak locates a symbol boundary in time and also SOF.  $M(n)$ is normally subjected to peak-finding, which is accepted provided the peak is above some threshold.   

Assuming high SNR, the samples are exactly repetitive, even with multipath, and $M(n)$ has maximum value 1.  However, the statistic is not guaranteed to be bounded by 1, and we have found the statistic above to be ill-behaved in transition between signal-present and signal-absent situations.  In particular, in a transition from signal-present to no signal, the denominator $R^2 (n)$ quickly drops to zero faster than the numerator, leading to high-amplitude peaks in $M(n)$ and thus possible false SOF declarations.  (Delaying $R(n)$ by $L$ samples to mitigate this proble, induces similar difficulties at the beginning
of a signal span.)

Though not mentioned in the Schmidl-Cox paper, the procedure is reminiscent of the Cauchy-Schwarz inequality for complex sums.  C-S holds that
\beq
|\sum_i a_i b_i |^2 \le \sum_i |a_i|^2 \sum_i |b_i|^2
\eeq
with equality iff $b_i=a_i^*$.  With a small change in the definition of $M(n)$ we can
claim the peak of $M(n)$ will never exceed 1, no matter the signal nature.  The proposed
modification is
\beq
\tilde M(n)=\frac{|P(n)|^2}{R (n)R(n-L)}
\eeq
i.e. we change only the denominator, and with no significant difference in computational
complexity.  By the C-S inequality, the maximum of $\tilde M(n)$ will be 1 for any signal and noise scenario, and
attain 1 when the first and second sets of $L$ samples are identical.  This provides a helpful self-scaling property, again holding in multipath conditiions.

The procedure can be made causal by allowing delay in the peak of $M(n)$ relative to the
start of the SOF symbol.  Defining $M'(n)=\tilde M (n-2L)$ gives
\beq
M'(n)=\frac{(\sum_{m=0}^{L-1} r^*_{n-2L+m} r_{n-L+m})^2}{\sum_{m=0}^{L-1}|r_{n-2L+m}|^2 \sum_{m=0}^{L-1}|r_{n-L+m}|^2}
\eeq
Now the `current' value of $M'(n)$ depends only on current and past inputs over a span of $2L$ samples.

A block diagram is sketched below.  The $L$-sample moving averages for $P$ and $R$ can be efficiently computed recursively if desired by an accumulator, adding a new sample and subtracting
the sample value $L$ samples earlier.

\begin{figure} [!hb]
	\centering
	\includegraphics[width=13cm]{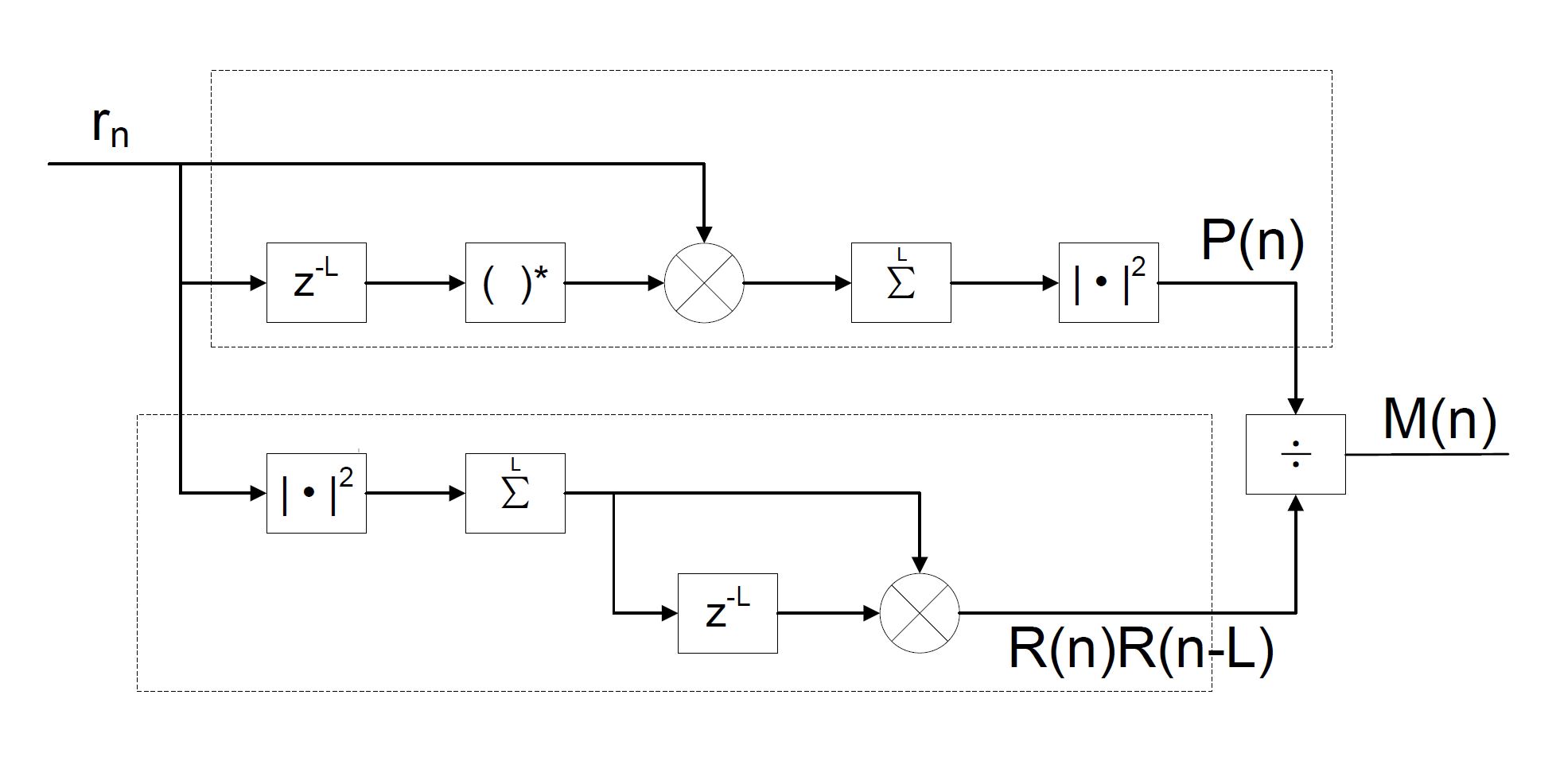}
	\caption{Block Diagram}
	\label{Block Diagram}
	
\end{figure} 

\section{Comparison}

We show simulation results for a case with $N=128$  (so $L=64$) on a Gaussian noise
channel with QPSK modulation and $E_b/N_0=10$ dB.  For illustration, we precede
a 16-symbol burst with the two-symbol Schmidl-Cox SOF preamble.  Additive noise precedes this
and also follows this interval in time.  In our study we have removed the cyclic prefix from
the first S-C symbol, eliminating the plateau in $M(n)$ without other consequence.  The expected peak in the SOF detector is at index 933 in subsequent plots.

Figure 2 shows the traces of $M' (n)$ for the conventional S-C detector, (Mold), as well as the proposed statistic (Mnew).  Note the rapid rise
in the ratio at the end of the 16-symbol interval for the S-C detector, in addition to the correct placement
of the SOF at the location of the S-C symbol.  On the other hand, the Mnew trace exhibits no such spurious peaks, while still correctly finding the correct SOF.

\begin{figure} [H]
	\centering
	\includegraphics[width=12cm]{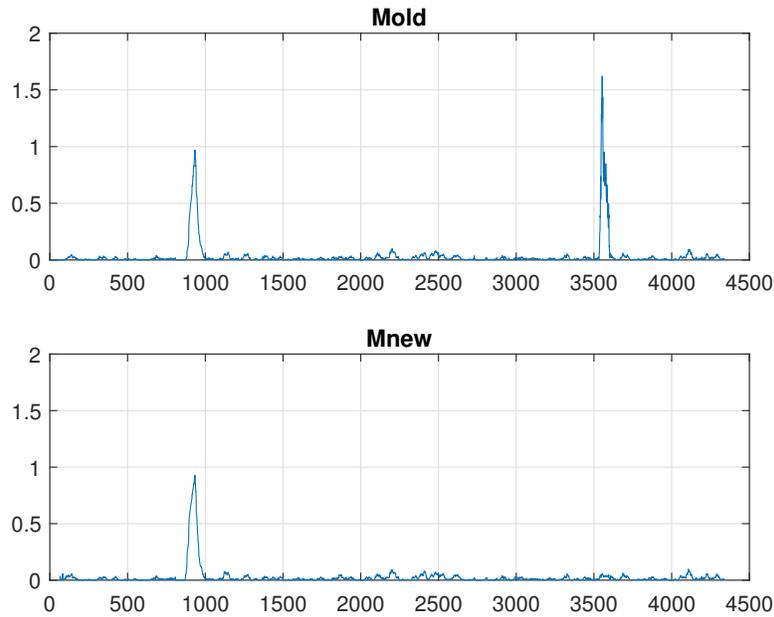}
	\caption{Comparison Between M-old and M-new, AWGN}
	\label{Comparisons}	
\end{figure}

A simple two-tap multipath test with multipath delay corresponding to a quarter symbol, and
tap weights $[0.8, 0.5 e^{j \pi /4}]$ was also done to confirm robustness of the procedure.   Figure \ref{multipath} below repeats the above traces, again at the same SNR.  Similar differences are noted, though here multipath actually reduces the spurious peak in Mold.

\begin{figure} [H]
	\centering
	\includegraphics[width=12cm]{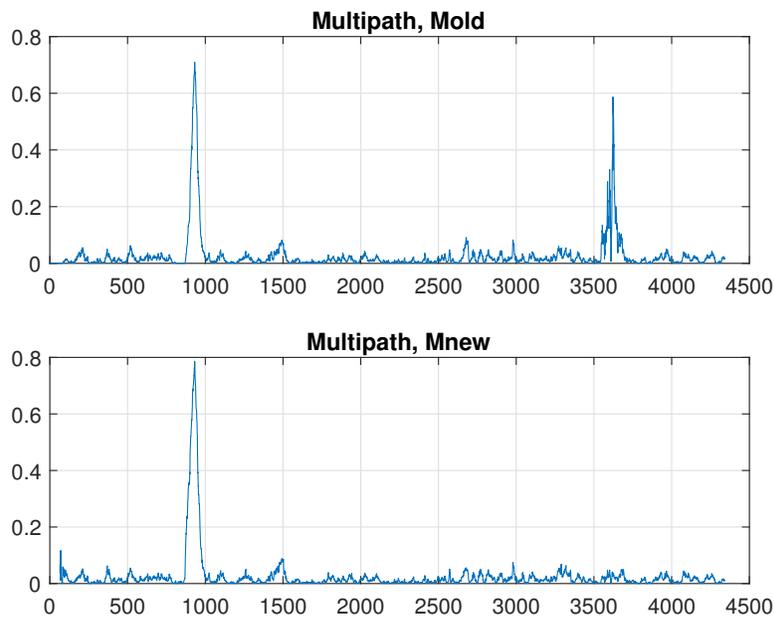}
	\caption{Multipath Comparison Between M-old and M-new}
	\label{multipath}
	
\end{figure} 
\subsection{Improved SNR at Peak}

The proposed timing detector also has better detection statistics at the proper peak of
the $M(n)$ signature, which we illustrate for a white Gaussian noise case with SNR=7 dB (energy per subcarrier divided by one-sided noise power density) 
by showing 
histograms of the original S-C statistic (Figure 4) and the modified one (Figure 5).  Clearly the proposed modification has a more concentrated p.d.f which translates to a better probability of detection versus probability of false alarm tradeoff.  The mean of the two statistics is the same; however the variance of the new statistic is smaller.
\begin{figure} [H]
	\centering
	\includegraphics[width=12cm]{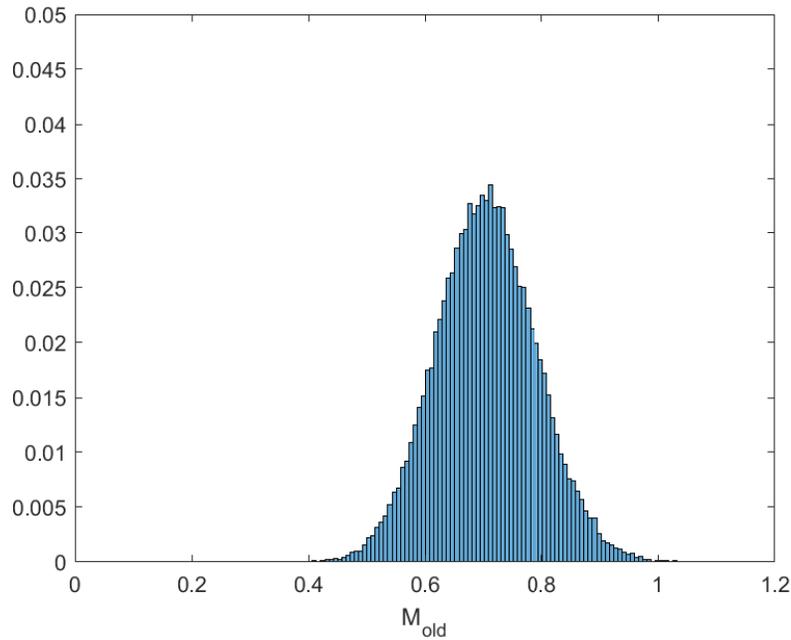}
	\caption{Normalized histogram for S-C statistic}
	\label{multipath}
	
\end{figure} 

\begin{figure} [H]
	\centering
	\includegraphics[width=12cm]{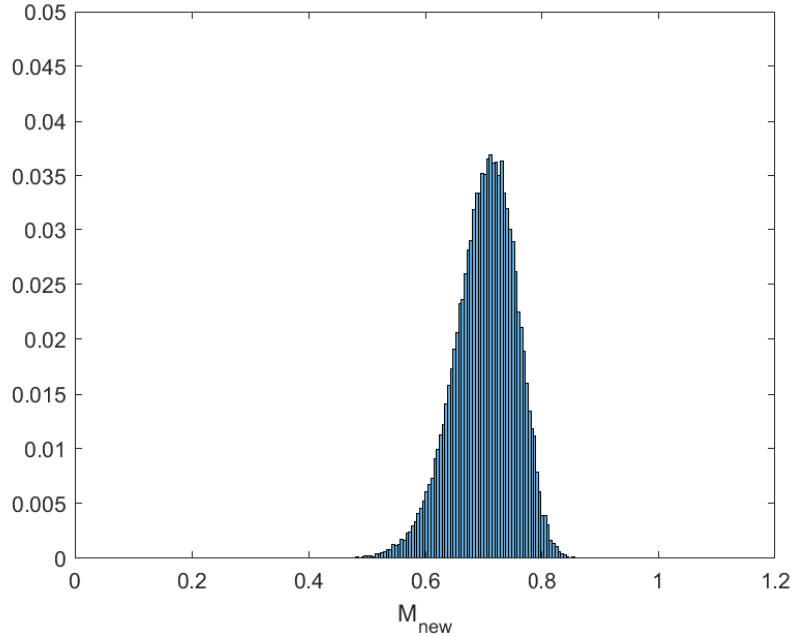}
	\caption{Normalized histogram for new statistic}
	\label{multipath}
	
\end{figure} 
An intuitive explanation for this result is that while the numerator statistic is identical in
both procedures, the modified denominator exhibits less variance as the product of two non-central chi-squared variates, versus the square of a single non-central chi-squared variate.
The latter involves a fourth moment of a complex Gaussian random variable, whereas the
modified denominator involves the product of two second moments.

\section{Conclusion}

A simple modification of the Schmidl-Cox detector for SOF in OFDM transmission is presented that eliminates false SOF's at beginning or end of transmission, while
simultaneously retaining an amplitude self-scaling property.  Moreover, the detector statistic
exhibits improved decision quality at the desired timing peak, and no increase in complexity relative to the original S-C algorithm is needed.

This discussion pertains to producing a clean, reliable SOF trigger signal.  A complete
detector needs a simple sliding peak-finding scheme, followed by a threshold test to
mitigate against false peaks in presence of noise alone.

Finally, the proposed modification is completely compatible with frequency estimation methods earlier proposed.

\end{document}